\documentclass[sigconf]{acmart}

\setcopyright{none}
\usepackage{dsfont}

\renewcommand\footnotetextcopyrightpermission[1]{}

\acmDOI{}
\acmISBN{}

\usepackage{amsmath}
\usepackage{mathtools}
\usepackage{booktabs}
\usepackage{graphicx}
\usepackage{microtype}
\usepackage{algorithm}
\usepackage{algpseudocode}
\usepackage{mathtools}
\mathtoolsset{showonlyrefs}

\newcommand{\R}{\mathbb{R}}
\newcommand{\E}{\mathbb{E}}
\newcommand{\Pc}{\mathcal{P}}

\newcommand{\Fc}{\mathcal{F}}
\newcommand{\Lc}{\mathcal{L}}

\renewcommand{\P}{\mathbb{P}}

\def\d{\mathrm{d}}

\def \Ac{{\mathcal{A}}}
\def \Bc{{\mathcal{B}}}

\def \Fc{{\mathcal{F}}}

\def \Lc{{\mathcal{L}}}
\def \Pc{{\mathcal{P}}}

\def \Wc{{\mathcal{W}}}

\newcommand{\CRPS}{\operatorname{CRPS}}

\begin{document}

\title{Deep-MKV-TS: Path-Dependent McKean--Vlasov Control for Financial
Time Series Generation \vspace{0.7cm}}

\author{Samer El Boustany}
\affiliation{%
  \institution{Murex SAS}
  \city{}
  \country{}
}
\email{selboustany@murex.com}

\author{Théo Basseras}
\affiliation{%
  \institution{Murex SAS}
  \city{}
  \country{}
}
\email{tbasseras@murex.com}

\author{Samy Mekkaoui}
\affiliation{%
  \institution{École Polytechnique}
  \city{}
  \country{}
}
\email{samy.mekkaoui@polytechnique.edu}

\author{Alexandre Alouadi}
\affiliation{%
  \institution{BNP-PAR and Ecole Polytechnique}
  \city{}
  \country{}
}
\email{alexandre.alouadi@polytechnique.edu}

\author{Yadh Hafsi}
\affiliation{%
  \institution{École Polytechnique}
  \city{}
  \country{}
}
\email{yadh.hafsi@polytechnique.edu}

\author{Huyên Pham}
\affiliation{%
  \institution{École Polytechnique}
  \city{}
  \country{}
}
\email{huyen.pham@polytechnique.edu}

\renewcommand{\shortauthors}{El Boustany et al.}

\begin{abstract} 
 We introduce Deep-MKV-TS, a path-dependent McKean–Vlasov framework for financial scenario generation. The stochastic dynamics are chosen by matching selected path and volatility features of generated scenarios to those observed in the data. Starting from an interpretable reference model, Deep-MKV-TS preserves the reference drift and adjusts its volatility, while a regularization penalty limits unnecessary departures from the calibrated dynamics. We solve the resulting control problem using a neural, sample-based implementation of the stochastic maximum principle.

We validate the method against an exactly computable oracle.
On Heston and Heston-mixture models, Deep-MKV-TS substantially
reduces path-dependent and volatility-related deficiencies of the
reference model. In delayed-volatility experiments, the correction
remains effective as the forecasting horizon increases, while direct
training becomes less reliable. On held-out intraday equity-index
futures, the corrected model improves conditional forecasts relative
to the reference and reaches a level of performance comparable to
flexible generative and historical baselines. The resulting scenarios
also support greater exposure than the reference under a fixed
drawdown-risk target. These results show that path-dependent
McKean--Vlasov control can enrich an interpretable reference model
without replacing it.

\end{abstract}



\maketitle

\fancyhead{}
\fancyfoot{}
\pagestyle{fancy}

\section{Introduction}
\label{sec:introduction}

Financial time series are central to many quantitative applications, including risk management, portfolio allocation, derivative pricing, stress testing, and optimal execution. They are also used to generate training data for data-driven trading strategies, such as reinforcement-learning methods \cite{Nevmyvaka2006,Espana2025QueueReactive, hafsi2026optimal}. In these settings, decisions depend on how prices and volatility evolve over time, not only on one-step returns. Realistic scenarios should therefore capture drawdowns, tail risk, and changing market conditions, making temporal dependence and persistent risk dynamics essential features of financial time-series models.

Financial decisions depend not only on individual returns but also on their path-dependent structure.
Two sets of scenarios can
have similar short-term returns and still imply very different risks if one
produces stronger volatility clustering, longer drawdowns, or a closer link
between early losses and future uncertainty \cite{cont2001empirical,bouchaud2009markets}. A useful
scenario generator must therefore reproduce both local price changes and the
way risk develops across an entire path.

Flexible generative models learn full path distributions directly and can be
powerful. They must, however, learn a high-dimensional path distribution from
finite data. Training can be demanding, and improving selected features
may come at the expense of dynamics that are already well calibrated.
 Because the complete mechanism is learned jointly, it can also be
difficult to control how the model changes and explain the resulting paths.

We study a complementary strategy. Rather than learning the complete
generation mechanism again, we retain a reference model and formulate a
path-dependent McKean--Vlasov control problem around it. Its drift remains
fixed, while volatility is adjusted to improve selected path features. A
penalty limits unnecessary changes to the reference. 
The correction is also traceable: at each time step, one can inspect
how much the controlled volatility departs from the reference and
which parts of the path trigger the adjustment.
The stochastic maximum
principle (SMP) provides the resolution method. This leads to the central question
of the paper: can a path-dependent McKean--Vlasov correction substantially
improve the path behavior of an imperfect financial model?

Our objective is therefore to measure how much a structured
path-dependent correction can improve a given reference model;
flexible generators serve only to contextualize the quality reached
after correction.

We call the method \emph{Deep-MKV-TS}: MKV refers to the McKean--Vlasov
dependence on the generative stochastic dynamics. The resulting forward-backward system arising from the SMP
traces these future effects back to the decision time, and a  neural
network learns the resulting correction from the history observed so far.
Training repeatedly generates scenarios, compares their path and volatility
features with the data, converts the remaining differences into correction
targets, and fits the network. 

The evidence is progressive. A small scenario tree provides a known optimum
for checking the exact and sampled calculations. A Heston mixture tests hidden and regime-dependent volatility. A delayed-volatility task
studies reliability as paths grow in number of steps. Finally, frozen models
for large-capitalization U.S. equity-index futures are evaluated
on unseen sessions, and a drawdown-risk decision.

The experiments support the correction at each level. Deep-MKV-TS
approaches the exact tree solution, substantially reduces the
reference errors on Heston and Heston-mixture data, and preserves
long-range dependence more reliably than direct training in the
delayed-volatility task. On held-out ES (S$\&$P 500), NQ (Nasdaq-100), and YM (Dow
Jones) sessions, it
improves all conditional-forecast scores over the reference and
reaches the range of flexible generative and historical baselines.
Its ES scenarios also support greater exposure under the prescribed
drawdown-risk limit.

\vspace{1mm}

\emph{\textbf{Our contributions}}. The paper makes three main contributions. First, we introduce a path-dependent McKean–Vlasov formulation for financial time-series generation in which a calibrated reference model is corrected rather than replaced. The controlled process preserves the reference drift, while its volatility is adjusted to reduce discrepancies in selected path and volatility features. A running penalty limits departures from the reference, while the
resulting volatility corrections remain directly traceable along
generated paths.

Second, we derive the associated stochastic maximum-principle conditions and turn the resulting forward–backward system into a practical learning algorithm. A recurrent neural network approximates the backward signal, allowing the volatility correction at each time to depend only on the observed history and to capture long-range path-dependent effects.

Third, we evaluate the method on both synthetic and real financial data. The experiments test hidden, regime-dependent, and delayed volatility effects, and assess performance on held-out intraday equity-index futures through conditional forecasts, and a downstream drawdown-risk decision. Across these settings, Deep-MKV-TS improves important path-dependent properties while preserving the structure of the reference model.

\vspace{1mm}

\emph{\textbf{Related works.}}
Financial path generation includes adversarial and signature-based models,
neural state-space models, neural stochastic differential equations, 
diffusion methods and optimal transport
\cite{wiese2020quant,ni2021sigwgan,zhou2023ls4,
kidger2021neuralsde,tashiro2021csdi,cao2026diffusion,rasul2021autoregressive,yoon2019time,alouadi2025robust,alouadi2026sbbts}. These flexible models learn a complete
generator directly and provide important comparisons for Deep-MKV-TS.
Relaxing the hard constraint in the optimal transport problem leads to a mean-field game or a mean-field control formulation; see, e.g., \cite{zhang2023mean,jiang2026schr}. The computational resolution of such formulations for generative modeling has recently been investigated in \cite{boustany2026learning} using SMP and Deep BSDE solver. 
Mean-field and path-dependent extensions allow the objective to depend on
the generated distribution and the complete history \cite{buckdahn2025path}.
Deep-MKV-TS combines these features in a discrete-time framework and
uses a neural network to estimate the conditional expectations arising
in the optimality system associated with the SMP. This leads to a
fully sample-based implementation that can be trained directly from
observed trajectories.

\section{A path-dependent generative model}
\label{sec:method}
\subsection{The reference model}\label{subsec : reference_model}

Deep-MKV-TS starts from an interpretable reference process providing
a conditional drift and volatility matrix. We use a multivariate
price-only extension of the path-dependent volatility model of
\cite{Guyon2023PathDependent}, in which past returns are summarized through short- and
long-memory trend and activity signals.

Let $
\Pi^{\rm obs}=\{t_1<\cdots<t_N=T\}$
and, for each path $p$, let $S^p_{t_i}\in\mathbb R_+^d$ for any $1 \leq i \leq N$. We work with
componentwise normalized log-prices and returns
\[
X^p_{t_i}
=
\log\!\left(\frac{S^p_{t_i}}{S^p_{t_1}}\right)\in\mathbb R^d,
\qquad
\Delta X_i^p=X^p_{t_{i+1}}-X^p_{t_i},
\qquad
\Delta t_i=t_{i+1}-t_i .
\]

We first fit the conditional mean of the next return. Given vectors $x_{t_1},\ldots, x_{t_N} \in \R^d$, we set $\boldsymbol{x}_{1:i} =(x_{t_1},\ldots, x_{t_i})$, $1 \leq i \leq N$. 
Let 
\begin{equation}\label{eq:qref}
    q_i^{\rm ref}(\boldsymbol{x}_{1:i})=\bigl(1,\psi_i^{\rm ref}(\boldsymbol{x}_{1:i})\bigr)
\end{equation}
where $\psi_i^{\mathrm{ref}}$ is a fixed (non-learned) map encoding the path
history through the short and long-memory trend signals of
\eqref{eq:trend-memory} below,
\begin{equation}\label{eq:embedding}
\psi_i^{\mathrm{ref}}(\boldsymbol{x}_{1:i})=\bigl(T_i^{(1)},T_i^{(2)}\bigr)
\in\R^{2d},
\end{equation}
each component standardized on the training set.
We set for any $1 \leq i \leq N$
\[
\widehat B_i
=
\operatorname*{arg\,min}_B
\sum_{p=1}^P
\left\|
\frac{\Delta X_i^p}{\Delta t_i}
-
B^\top q_i^{\rm ref}(X^p_{1:i})
\right\|^2
+
\rho_b\|B\|_F^2 ,
\]
where $\lVert \cdot \Vert_{F}$ denotes the Frobenius norm for matrices and $\rho_b >0$ is a ridge parameter.
The fitted conditional mean and residual are
\[
m_i^{\rm ref}(x_{1:i})
=
\widehat B_i^\top q_i^{\rm ref}(x_{1:i}),
\qquad
e_i^p
=
\Delta X_i^p
-
m_i^{\rm ref}(X^p_{1:i})\Delta t_i .
\]
The forward drift $b_i^{\rm ref}$ is either $m_i^{\rm ref}$ or zero,
as specified by the experiment.

For volatility, we maintain vector-valued trend and activity memories
$T_i^{(a)},V_i^{(a)}\in\mathbb R^d$, $a\in\{1,2\}$. Their short- and
long-memory components are combined as
\[
T_i=(1-\theta_T)T_i^{(1)}+\theta_TT_i^{(2)},
\qquad
V_i=(1-\theta_V)V_i^{(1)}+\theta_VV_i^{(2)} .
\]
The vector of marginal reference volatilities is
\[
v_i^{\rm ref}
=
\beta_0+\beta_1\odot T_i+\beta_2\odot\sqrt{V_i},
\]
where $\odot$ and the square root act componentwise. The memories are
updated componentwise according to
\begin{align}
T_{i+1}^{(a)}
&=
e^{-\lambda_{T,a}\Delta t_i}
\left(
T_i^{(a)}+\lambda_{T,a}\Delta X_i
\right),\label{eq:trend-memory}\\
V_{i+1}^{(a)}
&=
e^{-\lambda_{V,a}\Delta t_i}
\left(
V_i^{(a)}
+
\lambda_{V,a}
(v_i^{\rm ref})^{\odot 2}\Delta t_i
\right),\label{eq:activity-memory}
\end{align}
for $a\in\{1,2\}$. To account for contemporaneous dependence across coordinates, let
$R\in\mathbb S_{++}^d$ be a correlation matrix and define
\[
D_i^{\rm ref}
=
\operatorname{Diag}(v_i^{\rm ref}),
\qquad
C_i^{\rm ref}
=
D_i^{\rm ref} R D_i^{\rm ref}.
\]
The reference volatility matrix used in the dynamics is the symmetric
positive-definite square root
\[
\sigma_i^{\rm ref}
=
\left(C_i^{\rm ref}\right)^{1/2}
\in\mathbb S_{++}^d .
\]
Thus
$\sigma_i^{\rm ref}(\sigma_i^{\rm ref})^\top=C_i^{\rm ref}$.
The diagonal specification is recovered by taking $R=I_d$, and the
scalar model used in the experiments corresponds to $d=1$.

The parameters are estimated from returns by minimizing the Gaussian
negative log-likelihood
{\footnotesize\[
\mathcal L_{\rm ref}
=
\frac{1}{P(N-1)}
\sum_{p=1}^P\sum_{i=1}^{N-1}
\left[
\log\det \sigma_i^{{\rm ref},p}
+
\frac{1}{2\Delta t_i}
(e_i^p)^\top
\left(
\sigma_i^{{\rm ref},p}
(\sigma_i^{{\rm ref},p})^\top
\right)^{-1}
e_i^p
\right],
\]}
up to an additive constant. We fit the parameters on $80\%$ of the
training paths and select them on the remaining $20\%$. After
calibration, all reference parameters are frozen. Marginal
volatilities are clipped to $[\sigma_{\min},\sigma_{\max}]$ for
numerical stability where $\sigma_{\min}$ and $\sigma_{\max}$ are precised in Appendix \ref{sec:appendix-repro}.

\subsection{McKean-Vlasov problem formulation}


Let $M \geq N \in \mathbb{N}^{\star}$. Over $[0,T]$, we construct a finer grid $\Pi_{\text{disc}}$ 
\begin{align}
\Pi_{\text{disc}} &:= \big \lbrace (s_k)_{k \in \llbracket 0, M \rrbracket} : 0 \leq s_0 < s_1 < s_2 < \ldots < s_M =T \big \rbrace,
\end{align}
containing $\Pi_{\text{obs}}$, and used for the simulation of our generative stochastic dynamics in $\mathbb{R}^d$. We define
\begin{align}\label{eq : def_eta}
\eta(s) := \sup \big \lbrace i \in \llbracket 1, N\rrbracket : t_i \leq s \big \rbrace,
\qquad s  \in \Pi_{\text{disc}},
\end{align}
so that $t_{\eta(s)}$ is the largest observation point in $\Pi_{\text{obs}}$ not exceeding $s$. Given vectors $x_{t_1},\ldots, x_{t_N} \in \R^d$, we define $\boldsymbol{x}_{1:i} =(x_{t_1},\ldots, x_{t_i})$ for any $1 \leq i \leq N$ and we set $\Delta s_{k} = s_{k+1}- s_k$ for any $k \in \llbracket 0, M-1 \rrbracket$. 
On a filtered probability space $(\Omega,\mathcal{F}, \mathbb{P})$, we are given a sequence of independent standard gaussian random variables $(\epsilon_{s_{k}})_{k \in \llbracket 1, M \rrbracket}$.
We  denote  by $\mathbb{F}=(\Fc_{s_k})_{k \in \llbracket 1, M \rrbracket}$ the filtration generated by the sequence of random variables $(\epsilon_{s_k})_{k \in \llbracket 1, M \rrbracket}$, and eventually a uniform random variable used for randomization of the initial condition of the forward process $X$. we denoted  $\mathbb{E}_{s_k}[\,\cdot\,] := \mathbb{E}[\,\cdot\mid\mathcal{F}_{s_k}]$ for the conditional expectation given the $\sigma-$ algebra $\mathcal{F}_{s_k}$ and by $\mathrm{\mathcal{A}}$ the class of volatility controls valued in $\mathbb{S}^d_{++}$ the space of definite positive matrices of $\mathbb{R}^d$. Given $ \sigma \in \mathcal{A}$, our controlled state process $(X_{s_k})_{k \in \llbracket 0, M \rrbracket}$ is given by the following stochastic dynamics
\begin{align}\label{eq : state_dynamics}
X_{s_{k+1}} = & X_{s_k} + b^{\text{ref}}_{\eta(s_k)}(\boldsymbol{X}_{1:\eta(s_k)}) \Delta s_k  \\
& + \sigma_{s_k} \sqrt{\Delta s_k}\epsilon_{s_{k+1}}, ~ k \in \llbracket 0, M-1 \rrbracket ,
\end{align}
where the family of maps $(b^{\text{ref}}_i)_{1 \leq i \leq N-1}$ has been defined in Subsection \ref{subsec : reference_model}.
In the following, we denote $\mathbb{P}_{\boldsymbol{X}_{1:i}}$ the law of $\boldsymbol{X}_{1:i}$ under $\mathbb{P}$.
We now turn to the criterion to be optimized. Given some target path law $\nu \in \mathcal{P}_2 \big( (\mathbb{R}^d)^N \big)$ representing the joint distribution of normalized log-prices paths $ \big(\log(\frac{S_{t_{2}}}{S_{t_1}}), \ldots, \text{log}( \frac{S_{t_N}}{S_{t_{1}}}) \big)$, we aim to minimize the following cost functional
\begin{align}\label{eq: cost_functional}
    \Ac \ni \sigma \mapsto J(\sigma) = \quad & \E \Big[ \sum_{k=0}^{M-1} f_k( \boldsymbol{X}_{1:\eta(s_k)}, \sigma_{s_k}) \Delta s_k \Big] \\
   & +  G( \P_{\boldsymbol{X_{1:N}}}| \nu),
\end{align}

Here,  $G(\cdot|\nu) : \Pc_2 ( (\R^d)^N )   \to \R$ and $G(\mu | \nu)$ represents the discrepancy between $\mu$ and $\nu$, i.e. satisfies $G(\mu|\nu)\geq 0$ for any $\mu,\nu \in \Pc_2 \big( (\R^d)^N \big) $ with equality iff $\mu=\nu$.  In the following, we shall take 
\begin{align}\label{eq : def_discrepancy}
    G(\mathbb{P}_{\boldsymbol{X}_{1:N}}| \nu ) := & \quad   \frac{  w_{\text{path}}}{2} \text{MMD}^2( \P_{\boldsymbol{X}_{1:N}} | \nu) \\&  \quad + \frac{ w_{\text{vol}}}{2}  \text{MMD}^2(\varphi_{\text{vol}} \sharp \mathbb{P}_{\boldsymbol{X}_{1:N}}| \varphi_{\text{vol} } \sharp \nu),
\end{align}
for some penalty parameters $w_{\text{path}}$ and $w_{\text{vol}}$  and some map $\varphi_{\text{vol}}$ which  specifies the targeted stylized facts , namely we choose depending on the set of experiments, a customized vector of features $(\R^d)^N \ni \boldsymbol{x} \mapsto  \varphi_{\text{vol}}(\boldsymbol{x}) \in (\mathbb{R}^{d})^{N \times n_\text{features}}$, see Appendix~\ref{app:path-features} for more details. We also recall that $\sharp$ denotes the pushforward measure.
We recall that the maximum mean discrepancy (MMD) is defined as  
\begin{align}
    \text{MMD}^2(\rho_1|\rho_2) &= \E_{(U_1,U_2) \sim \rho_1^{\otimes^2} }[\kappa(U_1,U_2)] + \E_{(U_1,U_2) \sim \rho_2^{\otimes^2}}[\kappa(U_1,U_2)]  \notag \\
    &\quad - 2 \E_{(U,V) \sim \rho_1 \otimes \rho_2} [\kappa(U,V)],\notag
\end{align}
where $\kappa: (\R^d)^N \times (\R^d)^N \to \R$ is a symmetric positive kernel.
The use of the MMD is particularly convenient in our setting, as it
provides a sample-based discrepancy between probability distributions
that does not require estimating their densities. Moreover, when the
kernel $\kappa$ is characteristic (see \cite{sriperumbudur2010hilbert} for its rigorous definition), the MMD defines a metric on probability measures $\mathcal{P}_2 \big( ( \mathbb{R}^d)^N)$.

\vspace{1em}

The interpretation of \eqref{eq : def_discrepancy} is as follows: the first term
enforces the generated path distribution to match the target path distribution,
while the second term matches the distributions of the corresponding volatility
features through the pushforward map $\varphi_{\mathrm{vol}}$. The weights
$w_{\mathrm{path}},w_{\mathrm{vol}}>0$ balance the relative importance of these
two objectives. Moreover, for any $k \in \llbracket 0,M-1 \rrbracket$, the
running cost $f_k : (\R^d)^{\eta(s_k)} \times \Sigma$ is defined as
{\setlength{\abovedisplayskip}{3pt}
\setlength{\belowdisplayskip}{3pt}
\setlength{\abovedisplayshortskip}{3pt}
\setlength{\belowdisplayshortskip}{3pt}
\[
(\R^d)^{\eta(s_k)} \times \mathbb{S}^d_{++}
\ni (\boldsymbol{x},\sigma)
\mapsto
f_k(\boldsymbol{x},\sigma)
=
\mathcal{E}\!\left(
\sigma,\sigma^{\mathrm{ref}}_{\eta(s_k)}(\boldsymbol{x})
\right).
\]
}%
where the family of maps
$(\sigma_i^{\mathrm{ref}})_{1 \leq i \leq N-1}$
has been defined in Subsection~\ref{subsec : reference_model}
and where $\mathcal{E}$ is the specific entropy defined as
{\setlength{\abovedisplayskip}{3pt}
\setlength{\belowdisplayskip}{3pt}
\setlength{\abovedisplayshortskip}{3pt}
\setlength{\belowdisplayshortskip}{3pt}
\[
\mathcal{E}(\sigma,\bar{\sigma})
:=
\operatorname{Tr}\!\left(
\bar{\sigma}^{-1}(\sigma-\bar{\sigma})
\right)
-
\log\!\left(
\frac{\det(\sigma)}{\det(\bar{\sigma})}
\right),
\]
}%
for any $\sigma,\bar{\sigma}\in\mathbb{S}^d_{++}$.
\vspace{1 em}

The running cost $f_k$ penalizes deviations of the controlled volatility from
the prescribed reference volatility
$\sigma^{\mathrm{ref}}_{\eta(s_k)}(\boldsymbol{x})$. The deviation is measured
through the specific entropy, see \cite{benamou2024EST}. Geometrically, this divergence is naturally adapted to the space of
positive definite volatility matrices. It compares the controlled and
reference volatilities through their relative deformation, rather than
through an entrywise Euclidean distance. In particular, it accounts
jointly for changes in the principal directions and magnitudes of the
volatility matrices, while respecting their underlying matrix structure.

\subsection{The generative model and its associated optimal forward-backward system}
In order to solve the optimal control problem \eqref{eq : state_dynamics}-\eqref{eq: cost_functional}, we shall rely on a probabilistic method which has been extensively studied in the literature called Pontryagin's maximum principle, based on a calculus of variation,  see \cite{carmona2018probabilistic} . This allows to characterize the optimal control as the solution of a forward-backward system of stochastic equations which we shall derive in the following. Namely, for some small perturbation $\epsilon > 0$ of the control process $\sigma$ denoted by $ \sigma^{\epsilon}$, we are looking on a Gâteaux derivative of the cost functional $J$, namely for some limit, if it exists, of $\underset{\epsilon \to 0}{\text{ lim }} \frac{1}{\epsilon} \big( J(\sigma^{\epsilon})-J(\sigma) \big)$. Since, the cost functional $J$ depends on the law of the process $\boldsymbol{X}_{1:N}$, we need to introduce some calculus over the space $\Pc_2((\R^d)^N)$, and we decided to focus on the linear functional derivative. Formally, for any function $G : \Pc_2((\R^d)^m \to \R$ which admits a linear functional derivative, we denote by
$\frac{\delta}{\delta m}G(\mu)(\boldsymbol{x}) :
\Pc_2((\R^d)^N) \times (\R^d)^N \to \R$,
its linear functional derivative; see \cite{carmona2018probabilistic}
for a rigorous definition. It is simple to verify that
{\setlength{\abovedisplayskip}{3pt}
\setlength{\belowdisplayskip}{3pt}
\setlength{\abovedisplayshortskip}{3pt}
\setlength{\belowdisplayshortskip}{3pt}
\begin{align} \label{derivMMD} 
\partial_x \frac{\delta}{\delta m} \operatorname{MMD}^2(\mu,\nu)(\boldsymbol{x})
=
2\int_{(\R^d)^N}
\partial_x \kappa(\boldsymbol{x},\boldsymbol{y})\,(\mu-\nu)(\d \boldsymbol{y}).
\end{align}
}%


We now characterize the optimality system of a potential optimal
control.  We do not provide the full derivation of the associated system given below. We refer to \cite{Hafsi2026PathDependentSMP} and \cite{Mekkaoui2026DiscretePathMFC} for full details.

\begin{align}\label{eq : optimal_system}
\left\{
\begin{aligned}
&X_{s_{k+1}}^{\star}
=
X_{s_k}^{\star}
+
b_{\eta(s_k)}^{\mathrm{ref}}
\!\left(
\boldsymbol{X}^{\star}_{1:\eta(s_k)}
\right)
\Delta s_k
+
\sigma_{s_k}^{\star}
\sqrt{\Delta s_k}\,
\epsilon_{s_{k+1}},    \\[16pt]
& A^{\star}_{s_k} =\sum_{j=1, t_j \geq s_k}^{N} \bigg( \sum_{l=0, s_l \geq t_j}^{M-1} \Big(\partial_{x_j} f_{\eta(s_l)} (\boldsymbol{X}^{\star}_{1: \eta(s_l)}, \sigma^{\star}_{s_l}) \\ & \qquad \quad+ \partial_{x_j} b^{\text{ref}}_{\eta(s_l)}(\boldsymbol{X}^{\star}_{1: \eta(s_l)}) A^{\star}_{s_{l+1}} \Big)\Delta s_l  + \Gamma_j^{G}(\boldsymbol{X}^{\star}_{1:N}, \hat{\mu}^P) \bigg), \\[18pt]
&A^{\star}_{s_M} = \Gamma_N^{G}(\boldsymbol{X}^{\star}_{1:N}, \hat{\mu}^{P})
\\[18pt]
&Z_{s_k}^{\star}
=
\mathbb{E}_{s_k}
\!\left[
A_{s_{k+1}}^{\star}
\epsilon_{s_{k+1}}^{\top}
\right],
\\[16pt]
&\sigma_{s_k}^{\star}
= \Big( \big(\sigma^{\text{ref}}_{\eta(s_k)}\big)^{-1} 
+ \frac{1}{\sqrt{\Delta s_k}} Z_{s_k}^{\star} \Big)^{-1} 
\end{aligned}
\right.
\end{align}
for $k \in \llbracket 0, M-1 \rrbracket$,  and where we set 
\begin{align}\label{eq : derivatives_hamiltonian}
    \Gamma_{j}^G (\boldsymbol{x}, \mu):&= \partial_{x_{j}}  \frac{\delta}{\delta m}G(\mu|\nu)(\boldsymbol{x}), \\
    &=w_{\text{path}} \int_{(\R^d)^N} \partial_{x_j} k(\boldsymbol{x},\boldsymbol{y}) (\mu- \nu)(\d \boldsymbol{y}) \\
    &\quad + w_{\text{vol}} \int_{(\R^d)^N} \partial_{x_j} k(\varphi_{\text{vol}}(\boldsymbol{x}), \varphi_{\text{vol}}(\boldsymbol{y})) (\mu- \nu)(\d \boldsymbol{y}), 
\end{align}
for any $(\boldsymbol{x},\mu) \in (\mathbb{R}^d)^N \times \mathcal{P}_2((\R^d)^N)$ and any $j \in \llbracket 1, N \rrbracket$.

\subsection{Data sample-based implementation of Deep-MKV-TS}
\label{sec:implementation}

Starting from the forward--backward optimality system
\eqref{eq : optimal_system}, we aim to simulate the optimal forward process
$X^{\star}$. This requires the optimal controls
$(\sigma^{\star}_{s_k})_{k}$, hence the adjoint component
$(Z^{\star}_{s_k})_{k}$, which we approximate by a recurrent neural network
$\operatorname{GRU}_{\theta}$.
    Starting from the initial hidden state $h^{(m),p}_{s_{-1}}=0$ and initial
condition $X^{(m),p}_{s_0}=\xi$ for all $p\in\llbracket 1,P\rrbracket$,
where $\xi$ is a reference value or $0$, we iterate the following steps
for $m=0,\ldots,K-1$, with $K$ the total number of iterations.

\begin{enumerate}
\item[(1)] \emph{Forward simulation.}
For any path $p\in\llbracket 1,P\rrbracket$, we compute
\begin{align}
X_{s_{k+1}}^{(m),p}
&=
X_{s_k}^{(m),p}
+
b^{\mathrm{ref}}_{\eta(s_k)}
\big(\boldsymbol{X}^{(m),p}_{1:\eta(s_k)}\big)\Delta s_k
\notag\\
&\quad+
\sigma^{(m),p}_{s_k}\sqrt{\Delta s_k}\,
\epsilon_{s_{k+1}}^{(m),p},
\end{align}
where $\sigma_{s_k}^{(m),p}$ is obtained from the last equation of
\eqref{eq : optimal_system} evaluated at
$Z_{s_k}^{(m),p}=(\hat{Z}^{\theta^{(m)}})_{s_k}^{p}$, the output of a
recurrent network,
\begin{align}
\big((\hat{Z}^{\theta^{(m)}})_{s_k}^{p},\,h_{s_k}^{(m),p}\big)
=
\operatorname{GRU}_{\theta^{(m)}}
\big(X_{s_k}^{(m),p},\,h_{s_{k-1}}^{(m),p}\big),
\end{align}
whose hidden state $h_{s_k}\in\R^{96}$ is a 96-dimensional learned summary of the path up
to $s_k$ and is how the model represents memory.

\item[(2)] \emph{Ridge proxy.}
We approximate a proxy target
$(Z_{s_k}^{\mathrm{proxy},(m+1),p})_{k,p}$ by a ridge regression on the
realized adjoint observations
$A^{(m),p}_{s_{k+1}}\epsilon^{(m),p}_{s_{k+1}}$, where the pathwise adjoint
$A^{(m),p}$ is computed by automatic differentiation. Namely, we solve
\begin{align}
\beta^{(m+1)}_{s_k}
=
\underset{\beta}{\operatorname{arg\,min}}
\Big[\sum_{p=1}^{P}
\big|A^{(m),p}_{s_{k+1}}\epsilon^{(m),p}_{s_{k+1}}
-\beta^{\top}u_{s_k}^{(m),p}\big|^{2}
+\gamma|\beta|^{2}\Big],
\end{align}
where $u_{s_k}^{(m),p}=\Phi_k^{\mathrm{ref}}
\big(\boldsymbol{X}_{1:s_k}^{(m),p}\big)$, the map
$\Phi_k^{\mathrm{ref}}$ turning the simulated prefix into regression inputs
on the fine grid $\Pi_{\mathrm{disc}}$ exactly as $\psi_i^{\mathrm{ref}}$
does in \eqref{eq:embedding} on the observation grid
$\Pi_{\mathrm{obs}}$. Hence, we have
\begin{align}\label{eq: Z_proxy}
Z^{\mathrm{proxy},(m+1),p}_{s_k}
\approx
\big(\beta^{(m+1)}_{s_k}\big)^{\top}u_{s_k}^{(m),p}.
\end{align}

\item[(3)] \emph{Neural projection.}
We approximate $(Z^{(m+1),p}_{s_k})_{k,p}$ by minimizing the loss
\begin{align}
\Lc_{\mathrm{adj}}^{(m+1)}(\theta)
=
\frac{1}{PM}\sum_{p=1}^{P}\sum_{k=0}^{M-1}
\big\lVert(\hat{Z}^{\theta})_{s_k}^{p}
-Z_{s_k}^{\mathrm{proxy},(m+1),p}\big\rVert^{2}.
\end{align}
Then, we set $\theta^{(m+1)}
=
\underset{\theta\in\Theta}{\operatorname{arg\,min}}\;
\Lc_{\mathrm{adj}}^{(m+1)}(\theta)$ and iterate. 
\end{enumerate}

We then get the following algorithm for our implementation:

\begin{algorithm}[H]
\caption{Deep-MKV-TS training}
\label{alg:deep-mkv-training}
\footnotesize
\begin{algorithmic}[1]
\Require data paths $(\boldsymbol{\hat X}^{l}_{1:N})_{1\leq l\leq P}$;
frozen reference $(b^{\mathrm{ref}},\sigma^{\mathrm{ref}})$;
grid $(s_k)_{k=0}^{M}$; batch size $P$; iterations $K$;
ridge penalty $\gamma$
\Ensure trained parameters $\theta^{(K)}$
\State initialize $\theta^{(0)}$ with a zero final layer
\For{$m=0,\ldots,K-1$}
  \State set $h^{(m),p}_{s_{-1}}=0$,
    $X^{(m),p}_{s_0}=\xi$ for all $p\in\llbracket 1,P\rrbracket$, where $\xi$ a reference value or 0 
  \For{$k=0,\ldots,M-1$}
    \State $\big(\hat Z^{(m),p}_{s_k},h^{(m),p}_{s_k}\big)
      \gets\operatorname{GRU}_{\theta^{(m)}}
      \big(X^{(m),p}_{s_k},h^{(m),p}_{s_{k-1}}\big)$
    \State $Z^{(m),p}_{s_k}\gets\hat Z^{(m),p}_{s_k}$;
      \quad
      $\bar\sigma^{(m),p}_{k}\gets
      \sigma^{\mathrm{ref}}_{\eta(s_k)}
      \big(\boldsymbol X^{(m),p}_{1:\eta(s_k)}\big)$
    \State $\sigma^{(m),p}_{s_k}\gets
      \Big(
    \big(\bar{\sigma}_{k}^{(m),p}\big)^{-1}
    +\frac{1}{\sqrt{\Delta s_k}} Z_{s_k}^{(m),p}
\Big)^{-1}$
    \State draw $\epsilon^{(m),p}_{s_{k+1}}\sim\mathcal N(0,1)$
    \State $X^{(m),p}_{s_{k+1}}\gets X^{(m),p}_{s_k}
      +b^{\mathrm{ref}}_{\eta(s_k)}
      \big(\boldsymbol X^{(m),p}_{1:\eta(s_k)}\big)\Delta s_k$
    \Statex \hspace{3.75em}${}  \qquad \quad  +\sigma^{(m),p}_{s_k}
      \sqrt{\Delta s_k}\,\epsilon^{(m),p}_{s_{k+1}}$
  \EndFor
  \State We compute $(A^{(m),p}_{s_{k+1}})_{k=0,\ldots,M-1}$ with Eq. \eqref{eq : optimal_system},
  \For{$k=0,\ldots,M-1$}
    \State $u^{(m),p}_{s_k}\gets
      \Phi_k^{\mathrm{ref}}
      \big(\boldsymbol X^{(m),p}_{1:s_k}\big)$
    \State $\beta^{(m+1)}_{s_k}\gets
      \underset{\beta}{\operatorname{arg\,min}}
      \sum_{p=1}^{P}
      \big|A^{(m),p}_{s_{k+1}}\epsilon^{(m),p}_{s_{k+1}}
      -\beta^{\!\top}u^{(m),p}_{s_k}\big|^{2}
      +\gamma|\beta|^{2}$
    \State $Z^{\mathrm{proxy},(m+1),p}_{s_k}\gets
      \big(\beta^{(m+1)}_{s_k}\big)^{\!\top}
      u^{(m),p}_{s_k}$
  \EndFor
  \State detach the paths, replay the GRU, and set
  \Statex \hspace{3em}$\Lc_{\mathrm{adj}}(\theta)=
    \dfrac{1}{PM}\displaystyle
    \sum_{p=1}^{P}\sum_{k=0}^{M-1}
    \big\|\hat Z^{\theta,p}_{s_k}
    -Z^{\mathrm{proxy},(m+1),p}_{s_k}\big\|^{2}$
  \State $\theta^{(m+1)}\gets\operatorname{AdamW}
    \big(\theta^{(m)},
    \nabla_{\theta}\Lc_{\mathrm{adj}}\big)$
\EndFor
\State \Return $\theta^{(K)}$
\end{algorithmic}
\end{algorithm}

\section{Experiments}
\label{sec:experiments}

In the following experiments, we set $\Pi_{\text{obs}} = \Pi_{\text{disc}}$, i.e. $N=M$, and $d=1$.

\subsection{Benchmark generators and evaluation strategy}

Our primary comparison is between the frozen reference and its Deep-MKV-TS correction. SBTS 
\cite{hamdouche2026nonparametric,alouadi2025robust}, LS4 \cite{zhou2023ls4}, and CSDI \cite{tashiro2021csdi} provide external reference levels from distinct
families of flexible generators, with common data splits. Direct
training, obtained by a  direct network parametrization  and trained by minimization of the cost functional defined in \eqref{eq: cost_functional}, uses the same reference, network, objective, batch sizes,
and control bounds as Deep-MKV-TS, differing only in optimization. 

\subsection{Synthetic experiments}
\label{sec:synthetic}
We start with experiments where the true target is known exactly : 
a solvable tree, a Heston model, a mixture of Heston models and a delayed volatility experiment. 

We measure results with six scores, all lower-is-better: path distance is
the Sliced Wasserstein Distance (SWD) between standardized paths, averaged
over 256 random projections, realized-Volatility  error is the normalized $W_1$ (RV W1) distance
between realized-variance distributions, volatility clustering is the
error on the absolute-return autocorrelation ($|r|$ ACF), early--future error captures
how well the dependence between early history and later realized variance
is reproduced, regime error is the Total-Variation Distance (RTVD) between the
eight regime proportions, and maximum-drawdown error is the $W_1$ distance
between maximum-drawdown distributions (MDD $W_1$). The mathematical definitions of the metrics used are recalled in Appendix \ref{sec : evaluation-metrics} and the full architecture and training details are given in Appendix \ref{sec:appendix-repro}. 

\subsubsection{An exactly solvable tree}
We first test the full learning procedure in a small setting where the
correct answer can be computed exactly. We consider a GARCH time series with $N=6$. At each of six dates, the next
return can take only one of three possible values (up, down, neutral), giving $3^6 = 729$
complete paths. Since the probability of every branch is known, we can
enumerate all possible futures after each partial path instead of
estimating them from samples. This lets us compute both the exact
conditional signal $(Z_{s_k})_k$ and the best volatility correction at every
node of the tree.

We then hide these exact values from Deep-MKV-TS and run it following Algorithm \ref{alg:deep-mkv-training}. We set $P=128$ and $K=300$. For each prefix,
the model only sees $P$ sampled continuations, from which it estimate
the conditional expectation, store it in the GRU, and update the volatility
step by step. This learned procedure recovers 96.4\% of the objective
improvement achieved by the exact solution.

\subsubsection{Recovering Hidden Volatility from a Heston Model}
This experiment tests whether Deep-MKV-TS can recover hidden stochastic
volatility from prices alone. The target follows a Heston
process~\cite{heston1993closed}, but its instantaneous variance is never
revealed to the model. Paths are simulated under the Heston model with an initial variance of $0.04$, a long-run variance of $0.04$, a mean-reversion rate of $2$, a volatility of variance of $0.3$, and a price variance correlation of $-0.7$.
The reference model is fitted following the discussion in Subsection \ref{sec:method}.

Deep-MKV-TS substantially corrects the reference's volatility and
dependence errors (Table~\ref{tab:heston}): RV $W_1$ falls from
$0.089$ to $0.014$, $|r|$ ACF error from $0.068$ to $0.016$, and
early--future error from $0.214$ to $0.018$. The corrected model is
therefore in the same performance range as flexible generators while
retaining the reference dynamics.

\begin{table}[ht]
\caption{\textnormal{Heston: medians over four seeds, lower is better.}}
\label{tab:heston}
\centering\scriptsize
\setlength{\tabcolsep}{2pt}
\resizebox{\columnwidth}{!}{%
\begin{tabular}{lrrrrr}
\toprule
Method & SWD & RV $W_1$ & $|r|$ ACF & Early--future & MDD $W_1$\\
\midrule
Reference & 0.060 & 0.089 & 0.068 & 0.214 & 0.059\\
Deep-MKV-TS & 0.062 & 0.014 & 0.016 & 0.018 & 0.022\\
SBTS & 0.089 & \textbf{0.005} & \textbf{0.004} & \textbf{0.012} & 0.039\\
LS4 & \textbf{0.039} & 0.026 & 0.023 & 0.037 & \textbf{0.021}\\
CSDI & 0.104 & 0.204 & 0.011 & 0.080 & 0.118\\
\bottomrule
\end{tabular}}
\end{table}

\begin{figure}[t]
\centering
\includegraphics[width=\columnwidth]{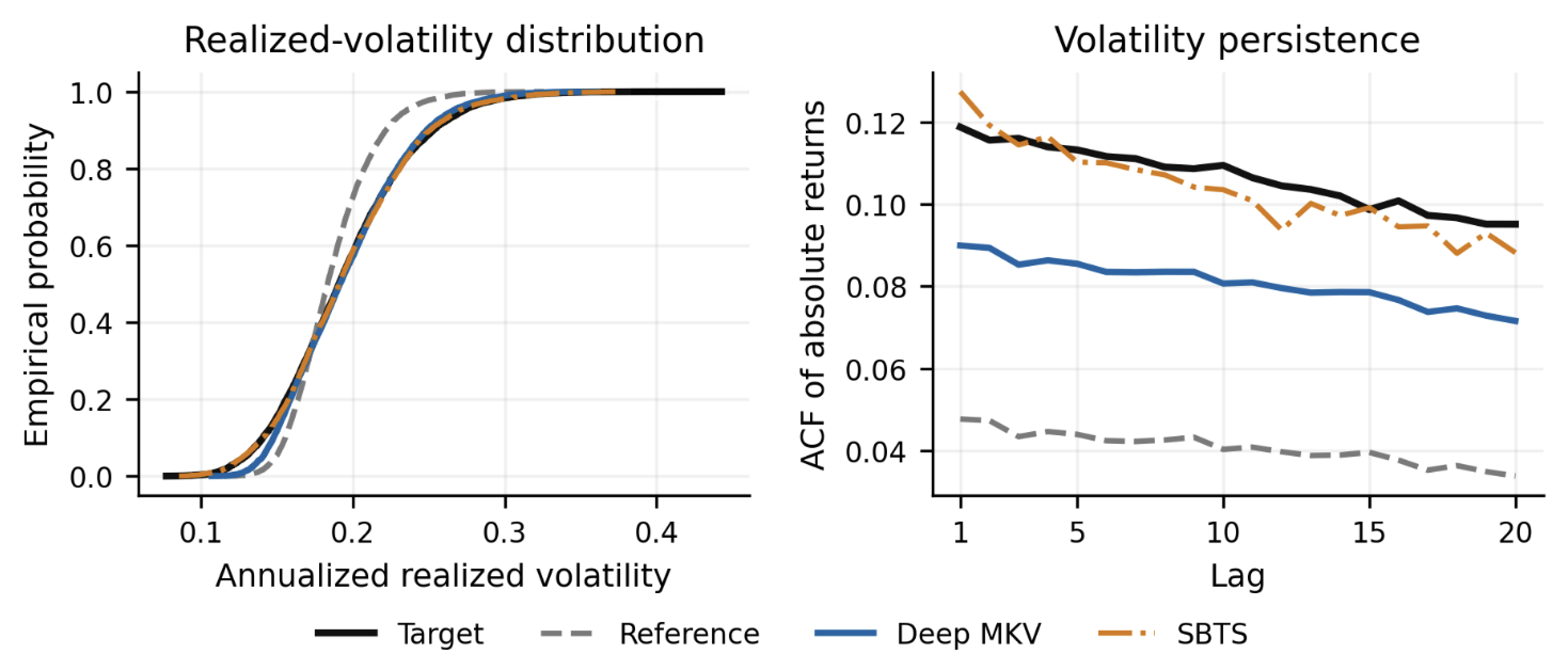}
\caption{\footnotesize{\textnormal{Comparison of volatility statistics for the target data, the Heston reference model, Deep MKV-TS, and SBTS. \textbf{Left:} Empirical cumulative distribution of annualized realized volatility. \textbf{Right:} ACF of absolute returns across different lags, measuring volatility persistence.}}}
\Description{Comparison of the distribution of realized volatility and the persistence of absolute returns for the target data, the Heston reference model, Deep MKV, and SBTS.}
\label{fig:heston_exp}
\end{figure}

\subsubsection{Recovering Hidden Volatility from a Mixture of Heston Models}

In this setting, we aim to recover the parameters of  a two-dimensional Heston model. Following the framework introduced in \cite{alouadi2025robust}, we sample eight
parameter sets $(\theta, \xi, \rho)$ uniformly from a given range, each
defining an equally likely regime where $\theta$, $\xi$ and $\rho$ represent respectively the mean-reversion parameter, the volatility of the variance and the correlation parameter. For every path, one of the eight regimes
is drawn and held fixed throughout its trajectory, with 1000 paths generated
per regime. The remaining parameters are shared across all regimes. 

Because the pooled reference blends the eight regimes, it provides a
deliberately misspecified starting point. Deep-MKV-TS substantially
reduces its errors across all diagnostics (Table~\ref{tab:mixture}),
bringing the corrected model to a level comparable to SBTS despite
retaining the reference structure.

\begin{table}[ht]
\caption{\textnormal{Mixture medians over four seeds, lower is better.}}
\label{tab:mixture}
\centering\scriptsize
\setlength{\tabcolsep}{2pt}
\begin{tabular}{lrrrrrr}
\toprule
Method & SWD & RV $W_1$ & $|r|$ ACF & Early--future & Regime TVD & MDD $W_1$\\
\midrule
Reference   & 0.347          & 0.479          & 0.322          & 0.765          & 0.629          & 0.309\\
Deep-MKV-TS & \textbf{0.237} & \textbf{0.095} & \textbf{0.046} & \textbf{0.046} & \textbf{0.331} & \textbf{0.009 }\\
SBTS        & 0.407          & 0.493          & 0.094          & 0.05 & 0.563          & 0.478\\
\bottomrule
\end{tabular}
\end{table}

\begin{figure}[t]
\centering
\includegraphics[width=0.5\columnwidth]{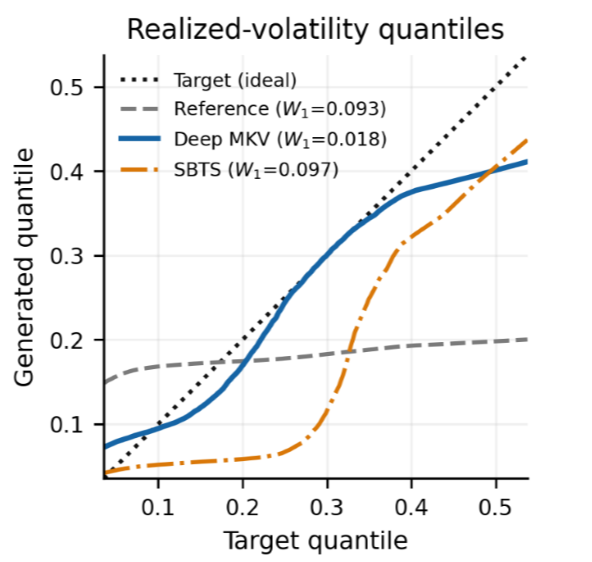}
\caption{\footnotesize{\textnormal{Quantile comparison of annualized realized volatility between the target data and the generated paths. The dotted line represents the ideal quantile matching, while the curves correspond to the Heston reference model, Deep MKV, and SBTS. The reported $W_1$ values quantify the discrepancy between the generated and target realized-volatility distributions.}}}
\Description{Quantile comparison of annualized realized volatility for the target data, the Heston reference model, Deep MKV, and SBTS.}
\label{fig:Mixture_Heston_exp}
\end{figure}

\subsubsection{Delayed volatility: recovering a long-memory effect}
This task tests a different kind of dependence than the previous two
experiments: not whether the generator recovers the overall volatility level
right, but whether it can remember an early event and correctly apply its
effect much later in the path. In the target process, a large drawdown
silently activates a memory that only increases volatility after a fixed
lag, then fades over time. 
The \emph{lagged-response error} measures how well the model reproduces the delayed effect of past drawdowns on future volatility, while the \emph{past--future error} measures how well it preserves dependence between early and later parts of the path. \emph{Deep dist.} and \emph{Direct dist.} report the percentage of the reference model's distributional error removed by Deep-MKV-TS and direct training, respectively. Similarly, \emph{Deep dep.} and \emph{Direct dep.} report the percentage of temporal-dependence error removed, with negative values indicating deterioration relative to the reference.
We test this at three horizons (128, 256, and 512 steps), scaling the
lag and decay proportionally, to see whether the correction still works
as the memory effect spans a longer history.

\begin{figure}[t]
\centering
\includegraphics[width=0.8\columnwidth]{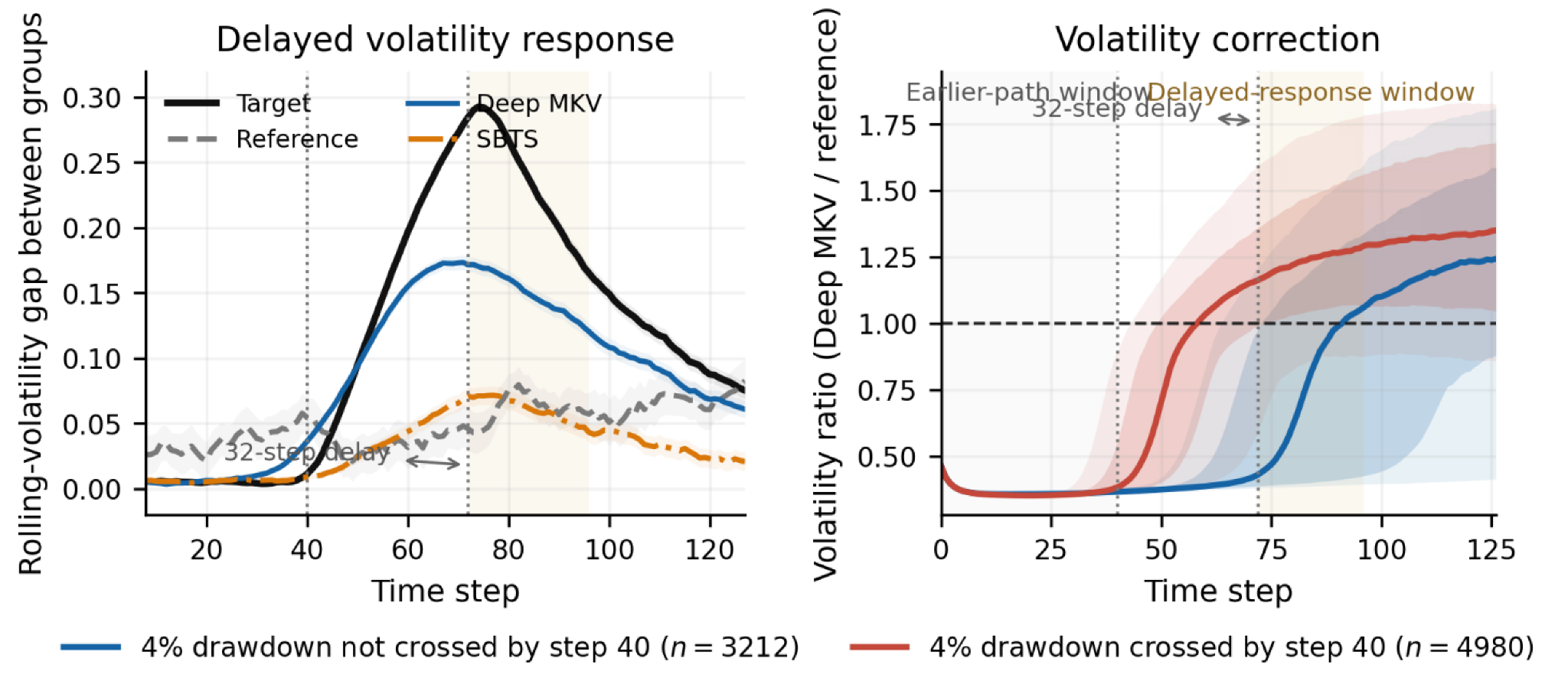}
\caption{\footnotesize{\textnormal{
Delayed-volatility experiment.
\textbf{Left:} Difference in rolling volatility between paths that
cross and do not cross the early drawdown threshold, showing the
delayed volatility response in the target and generated scenarios.
\textbf{Right:} Deep-MKV-TS volatility relative to the reference,
conditioned on the same two path groups. The learned correction
responds differently after the early-path event, making the timing
and magnitude of the path-dependent intervention directly traceable.
}}}
\Description{Comparison of the delayed-volatility response and recovered temporal dependence for the target, reference model, Deep MKV, and SBTS.}
\label{fig:Delayed_Volatility_exp}
\end{figure}

The correction strengthens with horizon, removing $31.6\%$ of the
reference distributional error at $N=128$ and $62.2\%$ at $N=512$
(Table~\ref{tab:delayed-merged}). At $N=512$, direct training gives lower RV $W_1$
($0.046$ vs.\ $0.055$) on its completed runs, but only $2/4$ seeds
complete successfully and temporal dependence deteriorates relative
to the reference. Deep-MKV-TS completes all runs and continues to
improve the reference dependence measures.
Figure~\ref{fig:Delayed_Volatility_exp} exposes the learned intervention directly:
the timing and magnitude of the volatility correction depend on the
observed path history and can be traced relative to the reference.

\begin{table}[t]
\caption{\footnotesize{\textnormal{Lagged volatility: horizon scaling (top) and detailed errors
at $N=512$ (bottom). All configurations complete $4/4$ seeds except
direct training at $N=512$ ($2/4$). Statistics are computed over
completed runs.}}}
\label{tab:delayed-merged}
\centering\scriptsize
\setlength{\tabcolsep}{4pt}
\begin{tabular}{lrrrr}
\toprule
$N$ & Deep dist. & Deep dep. & Direct dist. & Direct dep.\\
\midrule
128 & 31.6\% & 4.8\%  & 15.7\%  & $-6.7$\%\\
256 & 39.7\% & 6.9\%  & 14.1\%  & $-3.4$\%\\
512 & 62.2\% & 3.0\%  & 52.6\%  & $-4.5$\%\\
\midrule
Method & RV $W_1$ & $|r|$ ACF & Lagged-response & Past--future error\\
\midrule
Reference       & 0.208 & 0.032 & 0.194 & 0.672\\
SBTS            & 0.088 & 0.023 & 0.205 & 0.818\\
Deep-MKV-TS        & 0.055 & \textbf{0.017} & \textbf{0.174} & \textbf{0.665}\\
Direct training & \textbf{0.046} & 0.038 & 0.212 & 0.855\\
\bottomrule
\end{tabular}
\end{table}
\subsection{Real futures data}
\label{sec:futures}

\paragraph{Data and split.}
Data and split. We obtain Databento MBP-1 (market-by-price, depth-one) records for ES, NQ, and YM futures and aggregate them to one-second price series using the same procedure for all instruments. We use a common split of 496 training sessions (through May 2025), 144 validation sessions (June–December 2025), and 123 held-out test sessions (January–June 2026). Each dataset is trained separately.
We use three futures on large-cap U.S. equity indices: ES
, NQ  and YM. Each series consists of
continuous one-second price records. Across the three indices, we use a
common split of 496 training samples (through May 2025), 144 validation
samples (June--December 2025), and 123 held-out test samples (January--
June 2026). Each dataset is trained separately.

\subsubsection{Conditional evaluation}
\label{sec:conditional}

We evaluate the forecasting power of our model using a path shadowing approach~\cite{morel2023pathshadowingmontecarlo}. Given the first 65 log-prices values of a time series in the test
set, we retrieve the 256 nearest generated histories from a pool of
8,192 simulated paths (fit on the train set), and use their next 32 returns as the predictive
distribution for that test sample. Similarity between two histories is measured by Euclidean distance over
four standardized feature blocks, computed on the training bank: the
last 32 returns, a subsampled path shape (up to 24 points), rolling
volatility over 5-, 10-, and 20-step windows, and absolute- and
squared-return autocorrelations at lags 1, 2, 5, and 10. As two purely
historical baselines, we also build two real-data-only banks of 8{,}192
paths: a moving-block bootstrap that rebuilds each session from blocks
of eight consecutive returns, each drawn from a random training day but
kept at its original intraday position, and a session bootstrap, where a
session is an entire training day.

We score each predictive distribution with the continuous ranked
probability score (CRPS),
\begin{equation}
\CRPS(F,y)=\int_{-\infty}^{\infty}
\big(F(z)-\mathbf{1}\{y\leq z\}\big)^2\,dz,
\label{eq:crps}
\end{equation}
applied to cumulative return, individual returns, and future realized
volatility (lower is better). Because the score is conditioned on the
retrieved histories rather than computed unconditionally, it tests
whether the generated continuations respond appropriately to the
specific past observed for each test sample, not just whether the
overall distribution is realistic.

\paragraph{Results.}

Table~\ref{tab:futures} reports conditional-forecast performance on
the 123-session chronological held-out test. Deep MKV improves all
three CRPS scores relative to its matched fitted-drift reference for
each of ES, NQ, and YM. Thus, across all nine index--target
combinations, the path-dependent correction improves the frozen
reference model.

The learned and historical generators provide external reference
levels for the magnitude of this correction. After correction,
Deep MKV reaches performance comparable to these flexible alternatives
across the forecasting targets, while preserving the structure of the
underlying reference model.

\begin{table}[t]
\caption{\footnotesize{\textnormal{
Conditional-forecast CRPS ($\times 1000$; lower is better) on the
123-session chronological held-out test. The primary comparison is
between Deep MKV and its matched fitted-drift reference. The remaining
methods provide external benchmark levels. Deep MKV, its reference,
and SBTS are reported as four-seed mean $\pm$ sample standard deviation.
}}}
\label{tab:futures}
\centering
\scriptsize
\setlength{\tabcolsep}{3.0pt}
\renewcommand{\arraystretch}{1.08}
\begin{tabular}{llrrr}
\toprule
Index & Method
& Cum. return & Increment & RV \\
\midrule

ES
& Reference
& $1.045 \pm 0.004$
& $0.276 \pm 0.0002$
& $0.897 \pm 0.005$ \\

& Deep MKV
& $1.003 \pm 0.006$
& $0.267 \pm 0.0003$
& $0.634 \pm 0.013$ \\

& SBTS
& $1.040 \pm 0.010$
& $0.279 \pm 0.0003$
& $0.623 \pm 0.003$ \\

& Block bootstrap
& $1.042$
& $0.272$
& $0.771$ \\

& Session bootstrap
& $1.066$
& $0.282$
& $0.620$ \\

\midrule

NQ
& Reference
& $1.383 \pm 0.011$
& $0.361 \pm 0.0005$
& $1.092 \pm 0.002$ \\

& Deep MKV
& $1.334 \pm 0.002$
& $0.352 \pm 0.0004$
& $0.746 \pm 0.009$ \\

& SBTS
& $1.507 \pm 0.008$
& $0.383 \pm 0.0007$
& $0.907 \pm 0.013$ \\

& Block bootstrap
& $1.385$
& $0.359$
& $0.990$ \\

& Session bootstrap
& $1.432$
& $0.370$
& $0.802$ \\

\midrule

YM
& Reference
& $1.030 \pm 0.004$
& $0.266 \pm 0.0002$
& $0.704 \pm 0.005$ \\

& Deep MKV
& $1.013 \pm 0.006$
& $0.263 \pm 0.0004$
& $0.574 \pm 0.010$ \\

& SBTS
& $1.117 \pm 0.004$
& $0.297 \pm 0.0014$
& $0.608 \pm 0.007$ \\

& Block bootstrap
& $1.028$
& $0.265$
& $0.643$ \\

& Session bootstrap
& $1.070$
& $0.278$
& $0.585$ \\

\bottomrule
\end{tabular}
\end{table}

\subsubsection{A downstream drawdown-risk decision}
\label{sec:risk}
We now turn to a concrete use case: does a better generator lead to a
better risk decision, not just better forecasts? For each test sample, we look up the 256 generated paths whose past
history is closest to the observed one (Section~\ref{sec:conditional}),
and use their future 32 steps as the set of possible continuations. We
then estimate the largest expected future price drop by taking the 90th
percentile of the largest drawdown across these 256 continuations,
$\widehat q_s$. This estimate sets how much
exposure a trader could safely take while keeping expected losses under a
fixed 1\% budget: exposure is capped at 5 and shrinks as $\widehat q_s$
grows,
\begin{equation}
\ell_s=\min\left\{5,\frac{0.01}{c\,\widehat q_s}\right\},
\qquad
\text{violation}_s=\mathbf{1}\{\ell_s d_s> 0.01\},
\label{eq:risk-rule}
\end{equation}
where $d_s$ is the drawdown that actually occurred, $c$ is a safety factor
calibrated on validation data, and a violation means the realized loss
exceeded the 1\% budget despite the chosen exposure. This measures risk
capacity rather than trading profit, as a better generator should support
more exposure while keeping violations near the prescribed 10\% rate.

Relative to the fitted reference, Deep-MKV-TS increases average
exposure from $1.85$ to $1.99$ while remaining below the prescribed
$10\%$ violation limit, with an observed violation rate of $8.13\%$
(Table~\ref{tab:risk}). This provides a downstream interpretation of
the statistical correction: improving the conditional path
distribution recovers additional risk capacity from the reference
model without increasing violations beyond the target.

The historical baselines attain similar violation rates and, in the
case of the session bootstrap, a similar average exposure. We
therefore interpret this experiment as evidence that the correction
improves the decision usefulness of the reference scenarios, rather
than as evidence of universal dominance over alternative scenario
generation procedures.

\begin{table}[ht]
\caption{\footnotesize{\textnormal{Frozen ES drawdown-risk decision: mean $\pm$ std over
four seeds where applicable. T he prescribed violation limit is 10\%.}}}
\label{tab:risk}
\centering\small
\setlength{\tabcolsep}{4pt}
\begin{tabular}{lrr}
\toprule
Method & Average exposure & Violations\\
\midrule
Deep-MKV-TS & $\mathbf{1.99\pm 0.06}$ & $8.13\%\pm 0.66\%$\\
Fitted reference & $1.85\pm 0.09$ & $7.11\%\pm1.02\%$\\
SBTS & 1.87 & 6.50\%\\
Block bootstrap & 1.84 & 8.13\%\\
Session bootstrap & 1.95 & 8.13\%\\
\bottomrule
\end{tabular}
\end{table}

 \section{Discussion and Conclusion}
 \label{sec:discussion}

Deep-MKV-TS improves a reference financial model rather than
replacing it: the drift remains fixed while path-dependent
McKean--Vlasov control corrects volatility features missing from the
reference model. The correction is particularly effective for persistent,
regime-dependent, and delayed volatility effects, and improves
conditional forecasts and downstream risk capacity on held-out
futures. Overall, the results show that a structured correction can
bring an interpretable reference model toward the quality of flexible
generators while remaining traceable, retaining the structure it already captures and its interpretability.

\section*{Acknowledgements}

This work is part of a joint collaboration between Murex SAS and CMAP, École Polytechnique.

 Samy Mekkaoui is supported by the S-G Chair "Risques Financiers", and the "Deep Finance and Statistics" Qube-RT Chair.
 
 Alexandre Alouadi is supported by a CIFRE collaboration between BNP-PAR and École Polytechnique.
 
 Yadh Hafsi acknowledges support from the Chaire Risque Financiers, Société Générale, at École Polytechnique, and from the Institut Europlace de Finance (IEF).
 
Huyên Pham is supported by
the Chair ``Risques Financiers" Soci\'et\'e G\'en\'erale, and by FiME,
Laboratoire de Finance des Marchés de l'Energie, and the ``Finance and Sustainable Development'' EDF -
CACIB Chair.

\appendix
\section{Definition of the path features for the map $\varphi_{\text{vol}}$}
\label{app:path-features}
Let $\boldsymbol{x} = (x_{t_1}, \ldots, x_{t_N}) \in (\mathbb{R}^d)^N$ be a normalized log-price path, with log returns
\begin{align}\label{eq : return_def}
  r_i = x_{t_{i+1}} - x_{t_i}, \qquad i \in \llbracket  1, N-1 \rrbracket.
\end{align}

The broad path features are the complete path, its terminal value, and its return sequence:
\begin{align}\label{eq : def_maps_phi}
\begin{cases}
  \phi_{\mathrm{path}}(\boldsymbol{x}) &= (x_{t_1}, \ldots, x_{t_N}), \\
  \phi_{\mathrm{terminal}}(\boldsymbol{x}) &= x_{t_N}, \\
  \phi_{\mathrm{return}}(\boldsymbol{x}) &= (r_i)_{i=1}^{N-1}.
\end{cases}
\end{align}

The volatility feature map $\varphi_{\mathrm{vol}}$ is built from the following quantities.

\smallskip\noindent\textbf{Absolute returns.} We use the differentiable approximation $a_i = \sqrt{r_i^2 + \varepsilon_a^2} - \varepsilon_a$, which reduces to $|r_i|$ when $\varepsilon_a = 0$. The feature is $\phi_{\mathrm{abs}}(\boldsymbol{x}) = (a_i)_{i=1}^{N-1}$.

\smallskip\noindent\textbf{Squared returns.}  We set $\phi_{\mathrm{sq}}(\boldsymbol{x}) = (r_i^2)_{i=1}^{N-1}$.

\smallskip\noindent\textbf{Instantaneous realized variance.} Setting $(q_i = r_i^2 / \Delta t_i)_{i=1}^{N-1}$, the feature is defined as $\phi_{\mathrm{instRV}}(\boldsymbol{x}) = (q_i)_{i=1}^{N-1}$.

\smallskip\noindent\textbf{Global realized variance.} We set
\begin{align}\label{eq: globalRV}  \phi_{\mathrm{globalRV}}(\boldsymbol{x}) = \frac{1}{N-1} \sum_{i=1}^{N-1} q_i.
\end{align}

\smallskip\noindent\textbf{Rolling realized volatility.} For a window of length $w \in \llbracket 1, N-1 \rrbracket$,
\[
  v_i^{(w)} = \left( \frac{1}{w} \sum_{j=i-w+1}^{i} r_j^2 + \varepsilon_v^2 \right)^{1/2}, \quad i \in \llbracket  w,  N-1 \rrbracket.
\]
with $\varepsilon_v$ a small numerical constant. The feature is $\phi_{\mathrm{rollRV}}^{(w)}(\boldsymbol{x}) = (v_i^{(w)})_{i=w}^{N-1}$.

\smallskip\noindent\textbf{State-variance pairs.} To capture how volatility varies with the current state,
\begin{align}
  \phi_{\mathrm{stateRV}}(\boldsymbol{x}) = \big( (x_{t_i}, q_i) \big)_{i=1}^{N-1}.
\end{align}

\smallskip\noindent\textbf{Volatility-persistence features.} For any $1 \leq i \leq N-1$ and let $u_i$ denote either $a_i$ or $r_i^2$, standardized as $\widetilde{u}_i = (u_i - \overline{u}) / s_u$, where $\overline{u}$ and $s_u$ are its pooled mean and standard deviation. For a lag $\ell \in \llbracket 0, N-2 \rrbracket$,
\[
  \phi_{\mathrm{lag}, \ell}^{u}(\boldsymbol{x}) = \big( \widetilde{u}_i \, \widetilde{u}_{i+\ell} \big)_{i=1}^{N-1-\ell}.
\]
These features capture the persistence of large absolute or squared returns across different lags.

The complete volatility map collects all of the above:
\begin{multline}
  \varphi_{\mathrm{vol}}(\boldsymbol{x}) = \Big(
    \phi_{\mathrm{abs}}, \,
    \phi_{\mathrm{sq}}, \,
    \phi_{\mathrm{instRV}}, \,
    \phi_{\mathrm{globalRV}}, \, \\
    \phi_{\mathrm{rollRV}}^{(w)}, \,
    \phi_{\mathrm{stateRV}}, \,
    \big(\phi_{\mathrm{lag}, \ell}^{a}\big)_{\ell \in \mathcal{L}}, \,
    \big(\phi_{\mathrm{lag}, \ell}^{r^2}\big)_{\ell \in \mathcal{L}}
  \Big)(\boldsymbol{x}),
\end{multline}
where $\mathcal{L}$ is the chosen set of persistence lags.

\section{Evaluation metrics}\label{sec : evaluation-metrics}

\subsection{Mathematical definitions}

Given $p \geq 1$ and $\mu,\nu \in \Pc_p\big( (\R^d)^N \big)$, the space of probability measures with $p-th$ order, we define the Wasserstein distance as
\begin{align}
    \Wc_p(\mu,\nu) = \Big(\underset{\pi \in \Pi(\mu,\nu)}{\text{ inf }} \int_{(\R^d)^N \times (\R^d)^N} |x-y|_2^p \pi(\d x, \d y) \Big)^{\frac{1}{p}},
\end{align}
where $\Pi(\mu,\nu)= \big \lbrace \pi \in \Pc\big( (\R^d)^N \times (\R^d)^N \big) : \text{pr}_1 \sharp \pi = \mu \text{ and } \text{pr}_2 \sharp \pi = \nu \big \rbrace$ , where $\text{pr}_i$ denotes the projection maps for $i \in \lbrace 1, 2 \rbrace$ and where $| x-y|_2$ denotes the Euclidean norm of $(\R^d)^N$.  The Sliced Wasserstein distance associated to $p$ ($\text{SWD}_p$) is given by
\begin{align}\label{eq : SWD_def}
    \text{SWD}_p(\mu,\nu) = \Big(\int_{\mathbb{S}^{dN-1}} \Wc_p^p(P^{\theta} \sharp \mu, P^{\theta} \sharp \nu) \d \sigma(\theta)\Big)^{\frac{1}{p}},
\end{align}
with $\mathbb{S}^{dN-1} := \big \lbrace \theta \in (\R^d)^N : \lVert \theta \rVert_2 =1 \big \rbrace$ the sphere, $P^{\theta}(x) = \theta \cdot x$ the coordinate of the projection of $x \in (\mathbb{R}^d)^N$ on the vector space $\text{Vect}(\theta)$ and $\sigma$ representing the uniform measure on $\mathbb{S}^{dN-1}$. We shall notice that $P^{\theta} \sharp \mu \in \Pc_p(\R)$ for any $\mu \in \Pc_p \big((\R^d)^N \big)$ and we recall  \cite[Theorem 2.9]{santambrogio2015optimal} that we can rewrite \eqref{eq : SWD_def} as
\begin{small}
\begin{align}\label{eq : sliced_wasserstein_distance}
    \text{SWD}_p(\mu,\nu) = \Big( \int_{\mathbb{S}^{dN-1}} \int_{0}^{1} |F_{P^{\theta} \sharp \mu}^{-1}(t) - F^{-1}_{P^{\theta} \sharp \nu}(t)|^p \d t  \d \sigma(\theta)\Big)^{\frac{1}{p}},
\end{align}
\end{small}
where $F_{\mu}$ denotes the cumulative distribution function associated to $\mu$ and $F^{-1}_{\mu}$ its generalized inverse.
\\
The total variation distance $(\text{TVD})$ between two Borel probability measures $\mu,\nu$ on a Polish space $E$ is defined as
\begin{align}
    d_{\text{TVD}}(\mu,\nu) = \underset{A \in \Bc(E)}{\text{ sup}} | \mu(A)- \nu(A)|,
\end{align}
where $\Bc(E)$ denotes the Borel sets of $E$.
When $E =\lbrace x_1,\ldots, x_n \rbrace$, then we have \cite[Proposition 4.2]{levin2026markov}, 
\begin{align}\label{eq : total_variation_form}
    d_{\text{TVD}}(\mu,\nu) = \frac{1}{2} \sum_{k=1}^{n} | \mu(\lbrace x_k \rbrace) - \nu(\lbrace x_k \rbrace)|.
\end{align}
\subsection{Experimental evaluation metrics}
Let $\boldsymbol{x}=(x_{t_1},\ldots, x_{t_N}) \in (\R^d)^N$ be a normalized log-price path. We recall the notation introduced in \eqref{eq : return_def} for the log-returns.

\smallskip\noindent\textbf{SWD.}
The Sliced Wasserstein distance between the generated path and the true path is computed by $\text{SWD}_1(\mu_{\text{data}}, \mu_{\text{gen}})$ by the formula \eqref{eq : sliced_wasserstein_distance} where we set $\mu_{\text{data}}$ for the law of the data samples, $\mu_{\text{gen}}$ for the law of the generated samples 

\smallskip\noindent\textbf{Maximum drawdown.} We define the maximum-drawdown for the log-prices as
\begin{align}\label{eq : maximum_drawdown_map}
    \text{MDD}(\boldsymbol{x}) := \underset{0 \leq n \leq N}{\text{max}} \big( \underset{0 \leq k \leq n}{\text{ max }} x_k - x_n \big).
\end{align}
We then set
\begin{align}
    \text{MDD $\Wc_1$} := \frac{\Wc_1(\text{MDD} \sharp \mu_{\text{data}}, \text{MDD} \sharp \mu_{\text{gen}})}{\E_{X \sim \mu_{\text{data}}}[\text{MDD}(X)]}
\end{align}
where the MDD map is defined in \eqref{eq : maximum_drawdown_map}.

\smallskip\noindent\textbf{Realized volatility.}
We set
\begin{align}
    {\text{RV } \Wc_1}:= \frac{ \Wc_1 ( \phi_{\text{globalRV}} \sharp \mu_{\text{data}},  \phi_{\text{globalRV}} \sharp \mu_{\text{gen}} )} {\E_{X \sim \mu_{\text{data}}}[\phi_{\text{globalRV}}(X)]},
\end{align}
where we defined $\varphi_{\text{globalRV}}$ in \eqref{eq: globalRV}.

\smallskip\noindent\textbf{$V(r)$ ACF.} Let $\Lc$ denotes the set of lags.  For any $l \in \Lc$ and any $\mu \in \Pc_2((\R^d)^N)$, we define for any map $V : \R \to \R$ the quantity
\begin{align}
    \rho_{V}^{\mu}(l) := \frac{1}{N-1-l} \sum_{i=1}^{N-1-l} \text{Corr}_{\mu}(V(r_i), V(r_{i+l})).
\end{align}
Here, for any measurable functions $\Phi_1$ and $\Phi_2$,
$\operatorname{Corr}_{\mu}(\Phi_1,\Phi_2)$ denotes the correlation between
the random variables $\Phi_1(X)$ and $\Phi_2(X)$ when $X\sim\mu$.

We then define the corresponding discrepancy metric as
\begin{align}\label{eq : V ACF}
    \text{$V(r)$ ACF} := \big( \frac{1}{|\Lc|} \sum_{l \in \Lc} \big(\rho_{V(r)}^{\mu_{\text{data}}}(l) - \rho_{V(r)}^{\mu_{\text{gen}}}(l) \big)^2 \big)^{\frac{1}{2}},
\end{align}
Setting $V(r):=|r|$, we get the metric $|r|$ ACF.

\smallskip\noindent\textbf{Early-future.}
The early future metric is defined for the Heston and the mixture of Heston experiments as follows. Let
\begin{align}
    V_{\text{early}}(\boldsymbol{x}) := \big( \sum_{i=1}^{64} r_i^2 \big)^{\frac{1}{2}}, \quad V_{\text{future}}(\boldsymbol{x}) := \big( \sum_{i=65}^{96} r_i^2 \big)^{\frac{1}{2}}.
\end{align}
We then set
\begin{align}
   {\text{early-future}}= | \text{Corr}_{\mu_\text{data}}(V_{\text{early}}, V_{\text{future}}) - \text{Corr}_{\mu_\text{gen}}(V_{\text{early}},V_{\text{future}})|,
\end{align}
\smallskip\noindent\textbf{Lagged-response and past-future}
Let $h$ be the early-history cutoff  and $a > 0$ the prescribed threshold. We set the early-event indicator as
\begin{align}
E(\boldsymbol{x}):= \mathds{1}_{\underset{0 \leq n \leq h}{\text{max}} (\underset{0 \leq k \leq n}{\text{ max }} x_k - x_n) \geq a}.
\end{align}
Let $s > h$ denoting the beginning of future window and $H$ its length. We then define the global realized variance during the period $\llbracket s, s+ H \rrbracket$ as
\begin{align}
    V_{\text{R}_s^{s+H}}(\boldsymbol{x}):=  \big( \frac{1}{H} \sum_{i=s+1}^{s + H} \frac{r_i^2}{\Delta t_i} \big)^{\frac{1}{2}}.
\end{align}
We then define
\begin{align}
    \Delta_{V}(\mu) := \E_{X \sim \mu}[V_{\text{R}_s^{s+H}}(X) |E(X) =1]  - \E_{X \sim \mu} [V_{\text{R}_{s}^{s+H}}(X)|E(X)=0],
\end{align}
representing the difference in average future realized volatility between
paths for which the event $E$ occurs and those for which it does not.
\begin{align}\label{eq : lagged_response_past_future_error}
\begin{cases}
    \text{Lagged-response} &:=| \Delta_V(\mu_{\text{data}}) - \Delta_V(\mu_{\text{gen}})|, \\
    \text{Past-future error} &:=|\text{Corr}_{ \mu_{\text{data}}}(E(X), V_{\text{R}_{s}^{s+H}}(X)) \\
    &\qquad - \text{Corr}_{ \mu_{\text{gen}}}(E(X), V_{\text{R}_{s}^{s+H}}(X))|
\end{cases}
\end{align}
\smallskip\noindent\textbf{Regime TVD.} The regime TVD is defined for the mixture of Heston experiment. Let $l \in \mathbb{N}^{\star}$ be the number of mixture of components and assume that a classifier map $c^{\star}: (\R^d)^N \to \llbracket 1, l \rrbracket$ has been trained to identify the mixture component. Then, we set
\begin{align}
    \text{Regime TVD}:= d_{TVD}( c^{\star} \sharp \mu_{\text{data}}, c^{\star} \sharp \mu_{\text{gen}}),
\end{align}
where $\text{Regime TVD}$ is computed following \eqref{eq : total_variation_form}.

\smallskip\noindent\textbf{Dist and dep.} For any metric $m$, we define respectively $e_m^{\text{ref}}$ and $e_m^{\text{model}}$ as the reference and the model errors. We then define
\begin{align}
    e_m = \frac{e_m^{\text{ref}} - e_m^{\text{model}}}{e_m^{\text{ref}}} = 1 - \frac{e_m^{\text{model}}}{e_m^{\text{ref}}}.
\end{align}
A positive value for $e_m$ means an improvement, while a negative one means deterioration. We then define the metrics as 
\begin{align}
\begin{cases}
    \text{Dist} &:=\text{Median}(e_{\text{SWD}}, e_{\text{MMD $\Wc_1$}}, e_{\text{RV } \Wc_1}, e_{\text{Ter } \Wc_1},e_{\text{return } \Wc_1}),  \\
    \text{Dep} &:=\text{Median}(e_{r \text{ ACF}}, e_{|r| \text{ ACF}}, e_{r^2 \text{ ACF}}, e_{\text{L-R}}, e_{\text{P-F}}),
\end{cases}
\end{align}
where we set 
\begin{align}
\begin{cases}
    {\text{Ter } \Wc_1}= \frac{ \Wc_1 ( \phi_{\text{terminal}} \sharp \mu_{\text{data}}),  \phi_{\text{terminal}} \sharp \mu_{\text{gen}} )} {\sqrt{ \mathbb{V}_{X \sim \mu_{\text{data}}}[\phi_{\text{terminal}}(X)]}}, \\
     {\text{return } \Wc_1}= \frac{ \Wc_1 ( \phi_{\text{return}} \sharp \mu_{\text{data}}),  \phi_{\text{return}} \sharp \mu_{\text{gen}} )} {\sqrt{ \mathbb{V}_{X \sim \mu_{\text{data}}}[\phi_{\text{return}}(X)]}},
\end{cases}
\end{align}
with the maps $\phi_{\text{terminal}}$ and $\phi_{\text{return}}$ being defined in \eqref{eq : def_maps_phi}. Moreover, $r$ ACF, $|r|$ ACF, $r^2$ ACF are computed from \eqref{eq : V ACF} choosing respectively the maps  $V(r) =r$, $V(r)=|r|$ and $V(r)=r^2$ and where $e_{\text{L-R}}$ and $e_{\text{P-F}}$ refer respectively to the metrics lagged-response and past-future error defined in \eqref{eq : lagged_response_past_future_error}.

\section{Training Configuration}
\label{sec:appendix-repro}

Every run uses four training seeds and 8192 paths per split. We report
the median over seeds unless stated otherwise.

Deep-MKV-TS uses a one-layer, 96-unit GRU trained with AdamW, and a batch size
of 256 for both generated and target paths. On the synthetic tasks, we use
ridge penalty $10^{-3}$, weight decay $10^{-5}$, and volatility bounds
$[10^{-3}, 0.6]$. 
All other settings are listed in
Table~\ref{tab:settings}.

\begin{table}[H]
\caption{\textnormal{Locked Deep-MKV-TS settings.}}
\label{tab:settings}
\centering\scriptsize
\setlength{\tabcolsep}{3pt}
\renewcommand{\arraystretch}{1.05}
\resizebox{\columnwidth}{!}{%
\begin{tabular}{lllll}
\toprule
 & Heston &  Mixture & Delayed vol. & Futures\\
\midrule
$N$ & 128 & 128 & 128/256/512 & 128\\
$\Delta t$ & 0.004 & 0.004 & 0.004/0.002/0.001 & one session\\
$(w_{\text{path}},w_{\text{vol}})$ & (50,100) & (25,50) & (25,50) & (25,50)\\
$K$ & 2,500 & 3,000 & 3,000/2,500/2,500 & 400\\
Learning rate & 0.002 & 0.0015 & 0.002 & 0.00025\\
Drift rule & fitted & zero & zero & zero\\
\bottomrule
\end{tabular}}
\end{table}

\bibliographystyle{plain}
\bibliography{biblioSBB}

\end{document}